# Comment on "Frequency shifts in NIST Cs primary frequency standards due to transverse rf field gradients"

Kurt Gibble
Department of Physics, The Pennsylvania State University, University Park, PA 16802

We discuss the theoretical treatment of the microwave lensing frequency shift of the NIST-F1 and F2 atomic fountain clocks by Ashby *et al*. [Phy. Rev. A. **91**, 033624 (2015)]. The shifts calculated by NIST are much smaller than the previously evaluated microwave lensing frequency shifts of other clocks contributing to International Atomic Time. We identify several fundamental problems in the NIST treatment and demonstrate that each significantly affects their results. We also show a smooth transition of microwave lensing frequency shifts to the photon recoil shift for large wave packets.

The microwave lensing frequency shift is an atom-interferometric frequency shift of microwave atomic clocks [1]. The shift arises because the spatial curvature of a clock's microwave field amplitude produces dipole forces that weakly focus and defocus the atomic wave packets, leading to a frequency shift. This frequency shift typically ranges from $\delta\nu/\nu = 6\times10^{-17}$ to $9\times10^{-17}$ and a number of accurate clocks that contribute to International Atomic Time correct for this frequency bias [2-4]. The recent treatment by Ashby *et al*. predicts microwave lensing frequency shifts of the NIST-F1 and F2 fountain frequency standards that are much smaller, $1.7\times10^{-17}$ [5], whereas the treatment used for other clocks estimates the shift of NIST-F2 to be $8.7\times10^{-17}$ [6]. Although NIST has written [7-9] that the results in [5] "significantly extended and corrected the ideas presented" in [1], here we identify shortcomings and several fundamental problems in [5]. The errors include treating a microwave field that has a significantly smaller curvature than the fields in the actual clock cavities, finding that the microwave lensing frequency shift goes to zero for small microwave amplitudes, predicting a non-sinusoidal Ramsey fringe, and neglecting the spatial curvature of the microwave field amplitude over the atomic wave packets. We explain each of these below and, in the process, elucidate a connection between microwave lensing frequency shifts and photon recoil shifts [10]. Each of these errors in [5] make its results significantly incorrect.

The first error stems from NIST treating the microwave field of a cylindrical Ramsey cavity with no holes [6]. The holes in the cavity endcaps are necessary for the atoms to pass through the cavity and they produce significant perturbations of the cavity field [6,11]. The spatial curvature of the amplitude of the microwave field leads to dipole forces that focus and defocus the atomic wave packets, producing the lensing frequency shift. Ashby *et al*. consider a microwave field $B_z \propto J_0(3.83\rho/R)$, giving a position-dependent Rabi pulse area $\phi = \hbar b\pi \propto b_0 J_0(3.83\rho/R)$ (e.g. eqs. (3) & (7) in [5]), where R is the 3 cm cavity radius. A direct integration of Maxwell's equations very generally gives a Rabi pulse area $\phi = \hbar b\pi \propto b_0 J_0(k\rho)$, independent of the cavity's geometry [2-4,6,11,12]. Thus, $J_0(k\rho)$ produces 2.3 times larger dipole forces than $J_0(3.83\rho/R)$ for R=3 cm, and a similarly larger microwave lensing frequency shift.

Second, the NIST treatment yields a microwave lensing frequency shift that goes to zero for weak microwave fields [7-9]. Although [5] omits a discussion of the microwave lensing shift for the "fundamental" limit of weak fields [9], [7] explains that [5] yields a frequency shift "just like the microwave leakage shift," proportional to the scaled microwave intensity $b_0^2$, or $\phi^2$, consistent with discussions in [8]. In contrast, a key point of [1,6,12] is that the microwave lensing frequency shift is non-zero in the limit of zero microwave amplitude, for example, similar to the behavior of well-established photon recoil shifts [10].

To address the disagreement between Refs. [7-9] and [1,6,12], here we present a simplified derivation of the microwave lensing frequency shift in the limit of weak microwave fields. It shows that the microwave lensing frequency shift smoothly transitions to the photon-recoil shift as atomic wave packets become large and that the shift is similarly independent of the field amplitude [1,6,12]. We consider the simple case of a small atom cloud on the fountain axis and a one dimensional standing wave $\cos(k_x x)$. Following [1,12], but using the bare atom basis $|g\rangle$ and $|e\rangle$, small microwave field amplitudes imply that we need to only consider single microwave photon-recoils $\pm k_x$, $\cos(k_x x) = \tfrac{1}{2}\exp(ik_x x) + \tfrac{1}{2}\exp(-ik_x x)$, which destructively interfere for small x to produce the lensing shift [1]. In momentum space, the detected excited-state wave function just after the second Ramsey pulse is

$$\psi_e(k, t_2 = t_1 + T) = -\tfrac{i}{2}\phi \exp\left[i\left(kx - \tfrac{\hbar}{2m}k^2 t_2 - \alpha/2\right)\right]$$
$$\left[\exp(-i\omega_R T + i\alpha)\cos(k_x x - k v_R T) + \cos(k_x x)\right], \text{ using}$$

the microwave photon recoil velocity $v_R = \hbar k_x/m$, the photon recoil shift $\omega_R = \hbar k_x^2/2m$, and $\alpha$ as the phase shift of the second Ramsey pulse, corresponding to $\chi$ in [12].

Constructing a wave packet by integrating over k with a weight $\exp(-k^2\sigma_x^2/2)$ gives $\psi_e(k,t_2) = \tilde{A}\phi \exp(-x^2/2\sigma_x\tilde{w}_2)$ [$\exp(-i\omega_R T\tilde{w}_1/\tilde{w}_2 + i\alpha)\cos(k_x x\tilde{w}_1/\tilde{w}_2) + \cos(k_x x)$], where the complex wave packet sizes are $\tilde{w}_{1,2} = i\hbar t_{1,2}/m\sigma_x + \sigma_x$ at the two Ramsey interactions. The excited state probability $|\psi_e(k,t_2)|^2$ yields a Ramsey fringe with a frequency shift:

$$\delta\omega = \frac{\phi^2 \int_{-a}^{a} \cos(k_x x) \sum_{\pm} \pm e^{-\frac{x^2 \pm v_R xT + \frac{1}{2}v_R^2 T^2}{w_2^2}} \sin\left(w_{12}^2 \frac{k_x x \pm \omega_R T}{w_2^2}\right) dx}{\phi^2 T \int_{-a}^{a} \cos(k_x x) \sum_{\pm} e^{-\frac{x^2 \pm v_R xT + \frac{1}{2}v_R^2 T^2}{w_2^2}} \cos\left(w_{12}^2 \frac{k_x x \pm \omega_R T}{w_2^2}\right) dx} \quad . \tag{1}$$

Here $w_2 = |\tilde{w}_2|$, $w_{12}^2 = \hbar^2 t_1 t_2/m^2\sigma_x^2 + \sigma_x^2$, and 2a is the width of the detection aperture. The numerator is the perturbation of the transition probability due to microwave lensing and the denominator is the Ramsey fringe amplitude, both of which go to zero as $\phi^2$ (or $b_0^2$), yielding a non-zero frequency shift in the limit of zero microwave amplitude [1,6,12], in contrast to discussions in [7-9]. Fig. 1 shows that (1) reproduces the usual recoil shift in the limit of large atomic wave packets, $\sigma_x \to \infty$ (and $a \to \infty$), as expected. Detecting the center of a wave packet ($a/w_2 \ll 1$) yields the microwave lensing shift $\omega_R w_{12}/w_2$, which becomes equal to the photon recoil shift when wave packets spread minimally during the fountain time, $w_{12} \approx w_2$, for $\sigma_x > 10\mu m$.

The third fundamental problem in [5] is that the microwave lensing frequency shift distorts the clock's Ramsey fringe to be non-sinusoidal. From Fig. 3 of [5], the frequency shift of the center of the Ramsey fringe for small phase modulations $\alpha$ (top of the Ramsey fringe) is significantly larger than the frequency shift for larger $\alpha$ (e.g. $\pi/2$, the middle of the Ramsey fringe, where clocks normally operate). This reveals a general inconsistency with the interference of wave packets in quantum mechanics. The explicit spin algebra in [12] gives the detected excited-state amplitude, $\langle e|\Psi(\vec{r})\rangle = -\frac{1}{2}[(\psi_{22'}-\psi_{11'})\cos(\frac{\alpha}{2})+i(\psi_{21'}-\psi_{12'})\sin(\frac{\alpha}{2})]$, and the resulting Ramsey-fringe transition probability is:

$$P = \tfrac{1}{4}\int |\psi_{21'}-\psi_{12'}|^2 + \left(|\psi_{22'}-\psi_{11'}|^2 + |\psi_{21'}-\psi_{12'}|^2\right)\cos^2\left(\tfrac{\alpha}{2}\right) + \text{Im}\left[(\psi_{22'}-\psi_{11'})(\psi_{21'}^*-\psi_{12'}^*)\right]\sin(\alpha)\,d\vec{r}$$

which is a pure sinusoid [12]. Here the complex $\psi_{nm'}$'s are dressed state wave functions that depend on $\vec{r}$, where $\psi_{12'}$ represents a wave packet in dressed state $|1\rangle$ in the first Ramsey interaction and dressed state $|2\rangle$ in the second, after the temporal phase of the microwave field has been shifted by $\alpha$ [12].

Another problematic result in [5] is a significant dependence of the microwave lensing frequency shift on the initial wave packet size. This occurs because NIST treats only the gradient of the dipole energy at the center of a wave packet and neglects its curvature across the wave packet [5]. Minimum-uncertainty initial wave-packets that yield the experimentally observed velocity distribution spread quickly after the atoms are launched and have a significant spatial extent at the first Ramsey interaction [1,2]. The dipole force therefore varies significantly across these wave packets, weakly focusing and defocusing the two dressed states [1,12].

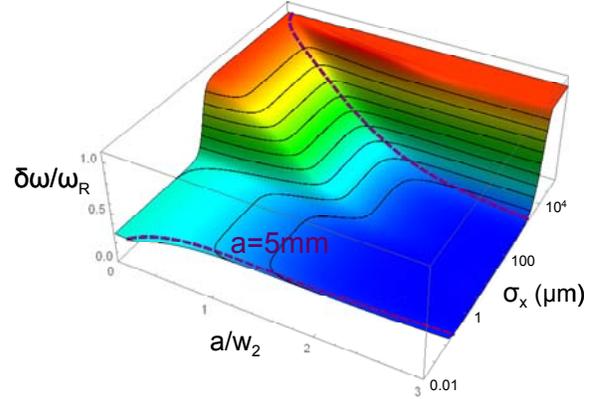

Fig. 1. (Color online) The microwave lensing shift (1) smoothly transitions to a photon recoil shift $\omega_R = \hbar k_x^2/2m$ for large initial wave packet sizes $\sigma_x$ and large apertures a. Here we consider cesium atoms and weak Ramsey pulses at $t_1$=0.21s and $t_2$=0.71s, with $k_x$=2π×9.192 GHz/c. The dashed lines depict an a=5mm aperture versus $\sigma_x$. For a small initial wave packet size $\sigma_x$, the quantum spread of the wave packet $w_2$ is large so the atoms are strongly clipped by the clock cavity aperture at $t_2$, yielding a microwave lensing frequency shift $\omega_R t_1/t_2$ [1,12]. As $\sigma_x$ grows, the spread $w_2$ decreases, yielding minimal clipping of atoms and no frequency shift [1,12]. If $\sigma_x$ would grow beyond a very large 5mm, the aperture would again clip the atom wave packets and the frequency shift goes to the recoil shift $\omega_R$.

Because NIST considers that the dipole force is everywhere zero for these atoms, they would incorrectly obtain no lensing frequency shift for a small, on-axis, initial atom cloud, such as in NPL-CsF2 [2]. This shortcoming of [5] necessarily implies that the microwave lensing frequency shift depends on the initial wave packet size (see Fig. 9 in [5]). Previously, [1] showed that the microwave lensing frequency shift depends on the velocity distribution of the cold atoms, but not on the initial minimum-uncertainty wave packet size. The lensing shift arises from the difference of the detected dressed state populations, "whether the dressed state populations are for one atom or the ensemble," which could have an incoherent thermal velocity distribution [1].

Unfortunately, it is difficult to pinpoint the error(s) in [5] that lead to the above incorrect dependence on microwave amplitude and to a non-sinusoidal Ramsey fringe since Ashby *et al.* do not provide concise or analytic expressions of their microwave lensing frequency shift. In contrast, using accurate approximations, we have previously derived analytic and explicit expressions for this shift [1,2,6,12]. These are highly useful for comparing and checking results, including numerical calculations.

We would like to directly address the stark statement by Ashby *et al.* that [1] "contains several results we find to be unphysical." However, neither do they justify this non-specific statement in [5] nor has a request yielded an explanation. Our response has therefore been to identify errors in [5,7] and provide additional physical descriptions of our results here and in [6,12].

The final issue is the importance of these errors for the accuracy of NIST-F2. We cannot address this with complete confidence because [5,7] do not contain the NIST-F2 fountain parameters that are needed to reproduce its results. Our unconfirmed estimates of their parameters yield a microwave lensing frequency shift of $\delta\nu/\nu = 8.7 \times 10^{-17}$ [6]. This is significantly larger than the prediction of $1.7 \times 10^{-17}$ in [5]. Our best estimate is also larger than the ascribed systematic error of $0 \pm 8 \times 10^{-17}$ for NIST-F2's combination of microwave lensing and other microwave amplitude dependent shifts, which is NIST-F2's largest systematic error [6,7]. In summary, the NIST treatment of the microwave lensing frequency shift [5] has several fundamental problems and each of these significantly affects their results.

We acknowledge financial support from NASA, the NSF, and Penn State.